\documentclass[apj,onecolumn]{emulateapj}
\usepackage{apjfonts}

\shorttitle{Angular momentum transfer in DM halos: erasing the cusp}
\shortauthors{Tonini et al.}
\journalinfo{ApJ accepted.}

\begin{document}

\title{Angular momentum transfer in \\ Dark Matter halos: erasing the cusp}

\author{C. Tonini\altaffilmark{1}, A.
Lapi\altaffilmark{1,2}, P. Salucci\altaffilmark{1}}

\altaffiltext{1}{Astrophysics Sector, SISSA/ISAS, Via Beirut 2-4,
I-34014 Trieste, Italy.} \altaffiltext{2}{Dip. Fisica, Universit\`a
``Tor Vergata'', Via Ricerca Scientifica 1, I-00133 Roma, Italy.}

\begin{abstract}
We propose that angular momentum transfer from the baryons to the
Dark Matter (DM) during the early stages of galaxy formation can
flatten the halo inner density profile and modify the halo dynamics.
We compute the phase-space distribution function of DM halos, that
corresponds to the density and anisotropy profiles obtained from
$N$-body simulations in the concordance cosmology. We then describe
an injection of angular momentum into the halo by modifying the
distribution function, and show that the system evolves into a new
equilibrium configuration; the latter features a constant central
density and a tangentially-dominated anisotropy profile in the inner
regions, while the structure is nearly unchanged beyond $10\%$ of
the virial radius. Then we propose a toy model to account for such a
halo evolution, based on the angular momentum exchange due to
dynamical friction; at the epoch of galaxy formation this is
efficiently exerted by the DM onto the gas clouds spiralling down
the potential well. The comparison between the angular momentum
profile gained by the halo through dynamical friction and that
provided by the perturbed distribution function reveals a surprising
similarity, hinting at the reliability of the process.
\end{abstract}


\section{Introduction}

The Dark Matter (DM) halos produced by numerical simulations feature
a cuspy density profile (Navarro et al. 1997, NFW) and a velocity
distribution isotropic in the center and slightly radially
anisotropic in the outer regions (Cole \& Lacey 1996, Thomas et al.
1998, Huss et al. 1999).

On the other hand, recent kinematical observations in galaxies are
at odds with the predicted DM distribution, favoring a cored profile
(Salucci \& Burkert 2000, Gentile et al. 2004). To solve this
striking disagreement, several authors (\textit{e.g.}, Spergel \&
Steinhardt 2000, Ahn \& Shapiro 2005) recurred to a framework
different from the standard cosmology (like, for instance,
self-interacting DM).

We adopt an evolutionary point of view to interpret the contradiction between
simulations and observations. We take for granted that a pure,
unperturbed DM halo is described by a cuspy NFW profile. We
recognize that the mentioned discrepancy exists precisely where the
baryonic component is present (see Donato et al. 2004); 
this leads us to question whether the baryons themselves could be at the origin
of the disagreement. Could a perturbation induced by the baryons affect
the dynamics and structure of the inner halo in such a way that the
original cuspy profile is modified in a corelike structure?

To test this possibility, we perform a specific study of the
dynamical microscopic properties of the DM halo through the
phase-space distribution function (DF), to conclude that 
a transfer of angular momentum is the main phenomenon at work 
in erasing the cusp. In fact, an angular momentum injection modifies both the
orbital energy of the DM particles and the velocity distribution
within the halo, breaking the symmetry of the system and leading to
a completely new equilibrium configuration.

Among the physical mechanisms viable to dynamically couple the 
baryons to the DM we follow the approach proposed by El-Zant et al. (2001, 2004),
according to which the interaction between the baryons and the halo is due to dynamical friction.
The latter is exerted by the DM on the clumpy material that falls into the center
of the halo during galaxy formation, and could be responsible for the
halo expansion and the transformation of the profile (for complementary analysis
see also Mo \& Mao 2004, and Navarro et al. 1996). 
In this framework, we specifically focus on angular momentum transfer,
and show that it can account for the halo 
evolution described by our treatment of the DF perturbation.

The plan of the paper is as follows: in Section 2 we compute the DF
that corresponds to the density and anisotropy profiles extracted from
$N-$body simulations; in Section 3 we perturb the DF with angular
momentum and show that a new equilibrium configuration is attained,
where the cusp has been smoothed out into a corelike feature and the
dynamics is dominated by tangential motions; in Section 4 we present
a physically-motivated toy model of angular momentum transfer
between the baryons and the DM halo, based on dynamical friction; in
Section (5) we summarize and discuss our findings.

\section{The NFW distribution function}

\begin{figure}\epsscale{0.75}\plotone{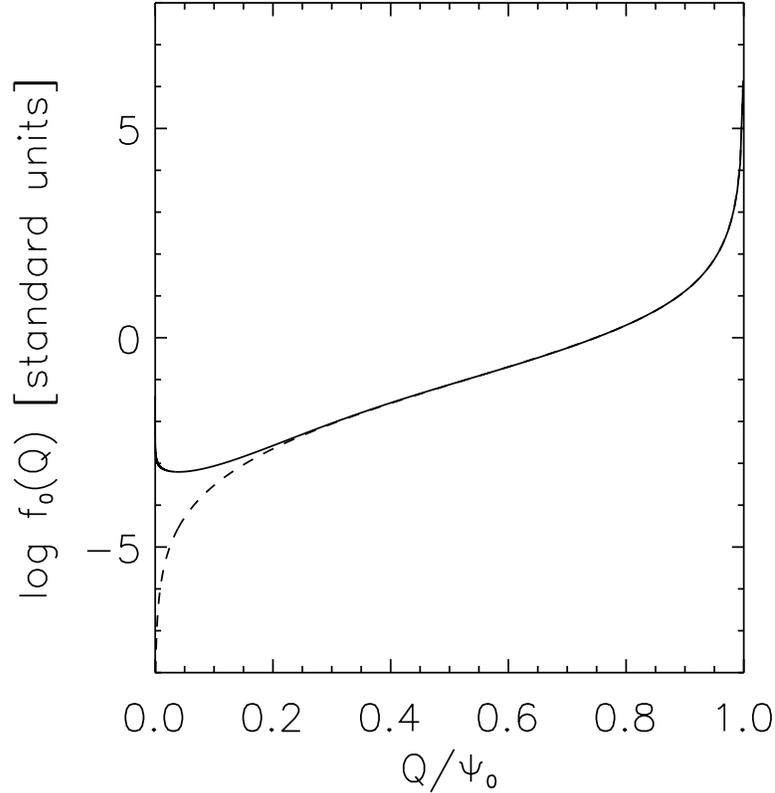}\caption{The phase-space
distribution function for a NFW halo, in standard units of $G = M_H
= R_H/2 = 1$, see Lokas \& Mamon (2001). \textit{Solid line}: halo
isotropic in the center and radially anisotropic at the outskirts;
\textit{dashed line}: totally isotropic.} \label{figDF}
\end{figure}

The hierarchical growth of the DM structures, as obtained from
$N-$body simulations, produces a density profile of the virialized
halos fitted by (Navarro et al. 1997)
\begin{equation}
\rho = {M_H\over 4\pi R_H^3}\, {c^2\, g(c)\over x\, (1+cx)^2}~;
\label{rhoNFW}
\end{equation}
here $M_H$ and $R_H$ are the virial mass and radius, $x=r/R_H$ is
the radial coordinate, $c$ is the concentration parameter, and
$g(c)\equiv [\ln{(1+c)}-c/(1+c)]^{-1}$. The NFW gravitational
potential is obtained through the Poisson equation $\nabla^2 \Phi =
4 \pi G \rho$ as
\begin{equation}
\Phi = -V_H^2\, g(c)\, \frac{\ln{(1+cx)}}{x}~, \label{phiNFW}
\end{equation}
where $V_H\equiv \sqrt{G\, M_H/R_H}$ is the virial velocity;
hereafter, we shall express all the physical quantities in terms of
$M_H$, $R_H$ and $V_H$.

The microscopic dynamical properties of the DM halo are linked to
its macroscopic density through
\begin{equation}
\rho(r) = \int f(r,v) \mathrm{d}^3 v~, \label{def}
\end{equation}
where $f$ is the halo phase-space DF, which is a function of all the
integrals of motions, and its symmetry properties determine those of
the system. Any macroscopic observable $O$ is obtained by means of
$f$ through the average $\langle O \rangle = \int O \, f \,
\mathrm{d}^3 v / \int f \, \mathrm{d}^3 v$.

In the simple case of a totally isotropic halo, $f$ is a function of
energy alone, uniquely determined from the density profile. By
conventionally defining the relative potential and binding energy as
$\Psi = -\phi$ and $\varepsilon = -E = \Psi-\frac{1}{2}v^2$
({\L}okas \& Mamon 2001), the DF is obtained from the
potential-density pair through the Eddington's inversion formula
(see Binney \& Tremaine 1987)
\begin{equation}
f(\varepsilon) = \frac{1}{\sqrt{8}\pi^2} \frac{d}{d \varepsilon}
\int_0^\varepsilon \frac{d \rho}{d \Psi} \frac{d
\Psi}{\sqrt{\varepsilon - \Psi}}~. \label{eddie}
\end{equation}
However, the simulated halos show a nontrivial anisotropy profile;
the degree of anisotropy is commonly expressed through the parameter
\begin{equation}
\beta(r)=1-\frac{\sigma^2_t}{\sigma^2_r},
\label{beta}
\end{equation}
where $\sigma^2_t$ and $\sigma^2_r$ are the tangential and radial
velocity dispersion respectively. In simulations the halos turn out
to be isotropic in the center ($\beta=0$), and radially anisotropic
($\beta > 0$) outwards.

For anisotropic spherical systems, $f$ is a function of both energy
and total angular momentum $L^2$, and there are infinite allowed DFs
that satisfy Eq.~(\ref{def}). We focus on a DF of the form
\begin{equation}
f(Q, L^2) = f_0\left(\varepsilon-{L^2\over 2 r_a^2}\right) \,
(L^2)^{\alpha}~, \label{fnostra}
\end{equation}
with $r_a$ being the anisotropy radius. The angular momentum
$\vec{L}(r)=\vec{r}\times\vec{v_T}$ is defined in terms of the
tangential velocity $\vec{v_T}(r)$; this is the $2D$ vectorial sum
(on spheres of radius $r$) of all the bulk velocities orthogonal to
the radial direction, and is a measure of the tangential component
of the internal, randomly-oriented motions of the halo. The orbital
energy $L^2/2r_a^2$ associated with $L$ lowers the binding
energy to give $Q=\varepsilon - L^2/2r_a^2$ (Osipkov 1979; Merritt
1985; Binney \& Tremaine 1987). In addition, the pure angular momentum component
of the DF is a simple functional form, a power-law of
index $\alpha$.

\begin{figure*}\plotone{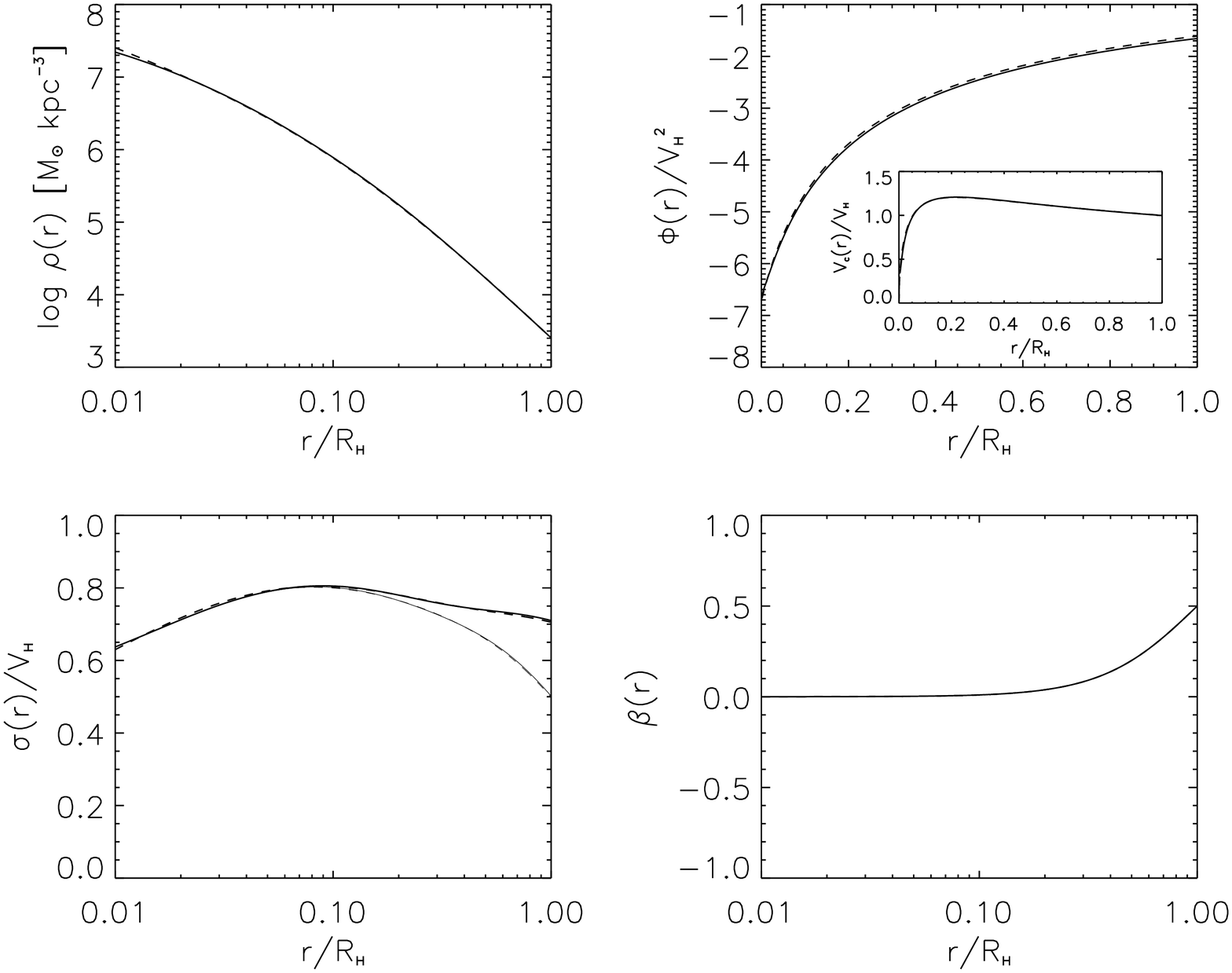}\caption{The reconstructed
NFW halo (\textit{solid lines}) compared to the original one
(\textit{dashed lines}). \textit{Upper left panel}: logarithmic
density profile; \textit{upper right}: gravitational potential and
rotation curve (\textit{inset}); \textit{lower left}: velocity
dispersions profiles, radial (\textit{thick}) and tangential
(\textit{thin}); \textit{lower right}: anisotropy profile.}
\label{figNFW}
\end{figure*}

By transforming the coordinate system from $(v_r,v_T)$ to $(Q,L^2)$
in Eq.~(\ref{def}), we can express the reconstructed density profile
as
\begin{equation}
\rho(r)=\frac{2 \pi}{r^2} \int_0^\Psi f_0(Q) \mathrm{d}Q \
\int_0^{2r^2 (\Psi-Q)/(1+r^2/r_a^2)} \frac{(L^2)^\alpha \ \mathrm{d}
L^2}{\sqrt{2(\Psi-Q)-(L^2/r^2) ( 1+r^2/r^2_a)}}~ \label{rhoNFWric}
\end{equation}
(for details see Binney \& Tremaine 1987, Chapters 4.4 - 4.5). In
spherical symmetry, the averaged $1D$-components of the velocity are
null, and the first non-zero moments are the radial and tangential
velocity dispersions
\begin{equation}
\sigma_r^2 (r) = \frac{2 \pi}{\rho r^2}\, \int_0^\Psi f_0(Q)
\mathrm{d}Q \, \int_0^{2r^2 (\Psi-Q)/(1+r^2/r_a^2)} (L^2)^\alpha \
\sqrt{2(\Psi-Q)-\frac{L^2}{r^2} \left(1+\frac{r^2}{r^2_a} \right)} \
\mathrm{d}L^2 ~, \label{sigmar}
\end{equation}
\begin{equation}
\sigma_t^2 (r) = \frac{\pi}{\rho r^4}\, \int_0^\Psi f_0(Q)
\mathrm{d}Q \, \int_0^{2r^2 (\Psi-Q)/(1+r^2/r_a^2)}
\frac{L^2\,(L^2)^{\alpha}}{\sqrt{2(\Psi-Q)-(L^2/r^2)(1+r^2/r^2_a)}}
\ \mathrm{d}L^2 ~, \label{sigmat}
\end{equation}
with the total velocity dispersion being
$\sigma^2(r)=\sigma^2_r(r)+2\sigma^2_t(r)$; thus the anisotropy
profile reads
\begin{equation}
\beta(r) = \frac{r^2-\alpha r^2_a}{r^2+r^2_a}~. \label{betar}
\end{equation}
Notice that, for positive $\alpha$, the anisotropy is tangential in
the inner regions where $r^2 < \alpha r^2_a$, zero at $r^2 = \alpha
r^2_a$ and radial in the outer regions where $r^2 > \alpha r^2_a$;
on the other hand, if $\alpha$ is negative the model is radially
anisotropic everywhere and for all values of $r_a$; therefore we
conclude that, with such a DF, $\alpha=0$ for the NFW simulated
halo. Moreover, we set the value of the anisotropy radius $r_a
\simeq 1$ ({\L}okas \& Mamon 2001). With this choice of $\alpha$ and
$r_a$ we obtain an isotropic inner region and a radially anisotropic
outer region.

Under these conditions, the relation between the DF and the density
profile reads (Cudderford 1991)
\begin{equation}
f_0(Q)= \frac{1}{2^{5/2} \pi^2} \frac{d^2}{dQ^2} \int_0^Q \left( 1 +
\frac{r^2}{r^2_a} \right) \rho(\Psi) \mathrm{d}\Psi~. \label{f0q}
\end{equation}
In Fig.~\ref{figDF} we plot the energy part of the DF defined by Eq.~(\ref{fnostra},
\ref{f0q}), for $\alpha=0$ and $r_a \simeq 1$ (\textit{solid line}),
compared with that of a totally isotropic halo (\textit{dashed
line}).

In Fig.~\ref{figNFW} we show the halo as reconstructed through the
DF defined by Eqs.~(\ref{fnostra}, \ref{f0q}) (\textit{solid
lines}), compared with the original NFW (\textit{dashed lines}); in
the upper panels we show the logarithmic density profile
(\textit{left}), and the gravitational potential (\textit{right}),
as well as the rotation curve (\textit{inset}). In the lower left
panel, we show the velocity dispersion profiles for the radial
(\textit{thick}) and the tangential (\textit{thin}) components; this
halo is isotropic in the inner $10\%$ of the virial radius, and
becomes radially anisotropic in the outer regions, as mirrored by
the anisotropy parameter profile (\textit{right}).

Once we know a suitable DF, we can investigate the specific
angular momentum profile, that is defined as follows:
\begin{equation}
\langle L(r) \rangle = \frac{2 \pi}{\rho r^2}\, \int_0^\Psi f_0(Q)
\mathrm{d}Q \, \int_0^{2r^2 (\Psi-Q)/(1+r^2/r_a^2)} \frac{L \
(L^2)^{\alpha}}{\sqrt{2(\Psi-Q)-(L^2/r^2)(1+r^2/r^2_a)}} \
\mathrm{d}L^2~. \label{jmedio}
\end{equation}
Notice that, even if the averaged $1D$ velocities are null, the
averaged angular momentum is nonzero, due to the symmetry of the DF,
as discussed after Eq.~(\ref{fnostra}). This is plotted as the
dashed line of Fig.~\ref{figAM}.

\section{Perturbing the halo: the angular momentum transfer}

From the above description of the halo, we can argue that the mass
distribution of the system is strictly linked to its dynamics. The
question that now arises is the following: is the halo stable
against perturbations in its dynamical state? In other words,
suppose that the halo becomes involved in a process that causes a
variation in its DF function, such as an increase of energy and
angular momentum; will the macroscopic observables, like the density
profile, the gravitational potential and the anisotropy profile, be
affected?

\begin{figure}\epsscale{0.75}\plotone{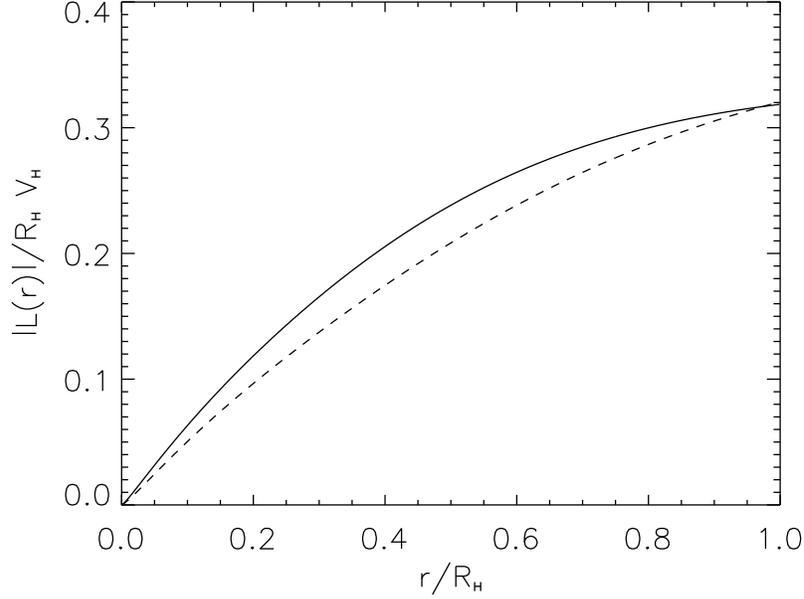}\caption{Angular momentum
profile yielded by the new phase-space DF (\textit{solid line}),
compared to the unperturbed NFW one (\textit{dashed line}), under
the same unevolved gravitational potential.} \label{figAM}
\end{figure}

We defer the reader to the next section for a toy model of angular
momentum transfer between baryons and DM during the first stages of
galaxy formation, and we proceed now to analyze its effects on the
equilibrium state of the halo.

Consider a system described by the DF of Eqs.~(\ref{fnostra},
\ref{f0q}), and suppose to inject angular momentum into it; this
translates into an increase of the particle orbital energy, that
becomes $L^2/2\, r_a^2+ \Delta(L^2/2\, r_a^2)$, and into a variation
of the power-law index $\alpha$. Since the DF is function of both
energy and angular momentum, the halo is bound to conserve $E$ and
$L^2$ before the perturbation, and $E+\Delta E$ and $L^2+\Delta L^2$
afterwise, redistributing the excess and rearranging the DM
particles in the $6D$-space of coordinates and velocities. This in
turn implies an evolution of the gravitational potential; hence, the
system moves towards a new equilibrium configuration of density and
velocity.

The process is governed by the Poisson equation
\begin{equation}
\frac{\mathrm{d}^2\Psi}{\mathrm{d}r^2} +
\frac{2}{r}\frac{\mathrm{d}\Psi}{\mathrm{d}r} = 4\pi\,G\,\int_0^\Psi
\int_0^{2r^2 (\Psi-Q)/(1+r^2/r_a^2)}\, f(Q,L^2)\,\mathrm{d}L^2\,
\mathrm{d}Q \label{poisson}
\end{equation}
where $\alpha$ and $r_a$ are now to be intended as the new,
perturbed parameters. This integro-differential equation has to be
solved for $\Psi$; the resulting potential, density and anisotropy
profiles are the observables of the new equilibrium state of the
halo. While the complete integration has to be done iteratively, the
solution for small radii is analytical, and gives an interesting
insight on the behavior of the halo.

Before proceeding, note that the density profile of
Eq.~(\ref{rhoNFWric}) can be written as
\begin{equation}
\rho(r)=\frac{(2\pi)^{3/2} \ 2^\alpha \
r^{2\alpha}}{(1+r^2/r_a^2)^{\alpha+1}}
\frac{\Gamma(\alpha+1)}{\Gamma(\alpha+3/2)} \ \int_0^\Psi f_0(Q)
(\Psi-Q)^{\alpha+1/2} \mathrm{d}Q~, \label{rhomezza}
\end{equation}
after the integration in $L^2$ is performed explicitly. For small
radii, \textit{i.e.} when $Q \rightarrow \varepsilon \rightarrow
\Psi_0$, with $\Psi_0$ the central value of the potential, it is
easy to see that $\rho(r) \propto 1/r \propto 1/[\Psi_0-\Psi(r)]$;
by changing variable in Eq.~(\ref{f0q}) from $\varepsilon$ to
$(\varepsilon-\Psi)/(\Psi_0-\varepsilon)$, the energy part of the DF
behaves like
\begin{equation}
f_0(\varepsilon) \propto (\Psi_0-\varepsilon)^{-5/2}~.
\label{f0aprox}
\end{equation}
Putting this expression into Eq.~(\ref{rhomezza}), and passing from
$\varepsilon$ to $(\Psi-\varepsilon)/(\Psi_0-\Psi)$, the density
profile at small radii is found to behave as
\begin{equation}
\rho(r) \propto [\Psi_0-\Psi(r)]^{\alpha-1} r^{2\alpha}~.
\label{rhosmall}
\end{equation}
This expression is now inserted into the Poisson equation
(\ref{poisson}), to obtain the self-consistent solution for the new
potential $\Psi(r)$, which reads
\begin{equation}
\Psi_0-\Psi(r) \propto r^{2\,(\alpha+1)/(2-\alpha)}~.
\label{psismall}
\end{equation}
Finally, the new density profile $\rho(r)$ from Eq.~(\ref{rhosmall})
reads
\begin{equation}
\rho(r) \propto r^{-2\,(1-2\alpha)/(2-\alpha)}~; \label{rhosmall2}
\end{equation}
thus we find that for $\alpha \rightarrow 0$ the inner profile behaves like
$r^{-1}$ (NFW), while for $\alpha \rightarrow 1/2$ we obtain $\rho(r)
\rightarrow \ constant$.

\begin{figure*}\plotone{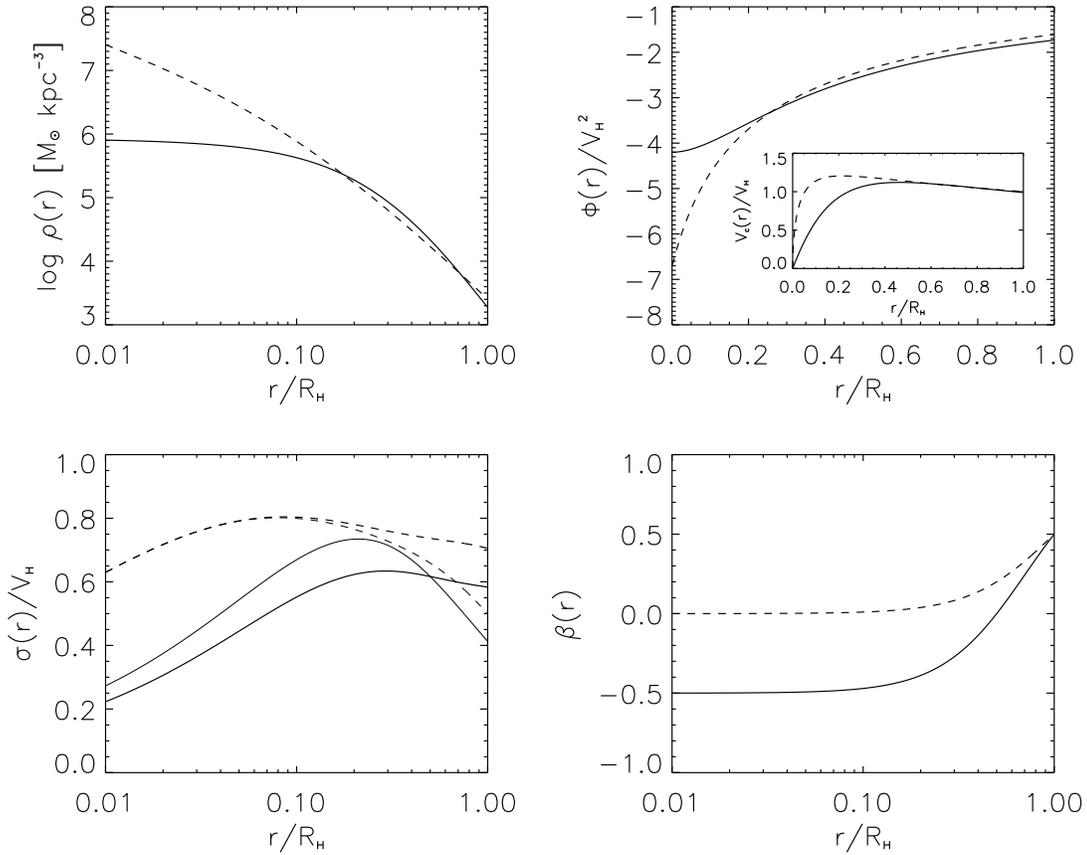}\caption{New equilibrium
configuration of the perturbed halo (\textit{solid lines}), compared
to the NFW (\textit{dashed lines}). \textit{Upper left panel}:
logarithmic density profile; \textit{upper right}: gravitational
potential and rotation curve (\textit{inset}); \textit{lower left}:
velocity dispersions profiles, radial (\textit{thick}) and
tangential (\textit{thin}); \textit{lower right}: anisotropy
profile.} \label{figNEW}
\end{figure*}

In addition, the anisotropy parameter defined by Eq.~(\ref{betar})
behaves like $\beta(r) \rightarrow -\alpha$, and for $\alpha=1/2$
the halo features a constant tangential anisotropy in the inner
regions. The inner specific angular momentum profile depends on the
value of $\alpha$ as well, and is obtained as
\begin{equation}
L(r) \propto r \ [\Psi_0-\Psi(r)]^{1/2} \propto r^{3/(2-\alpha)}~;
\label{jsmall}
\end{equation}
for the unperturbed halo $L(r) \rightarrow r^{3/2}$ and for
$\alpha=1/2$ we get $L(r) \rightarrow r^2$.

Beside these analytical results, we performed the full numerical
integration of Eq.~(\ref{poisson}). As a boundary condition
throughout this computation we adopted the halo mass conservation;
we further normalized the evolved potential in order to obtain the
same behavior of the outer rotation curve as before.

It is interesting to analyze this process in two steps: since the
density and anisotropy profiles and the potential well evolve
together, it is impossible to evaluate the variation in the angular
momentum of the system after the new equilibrium state is reached,
because the system has then lost memory of its initial conditions.
Instead, suppose to picture the halo at the moment when the DF
received its input in energy and angular momentum, but the potential
well has not yet evolved; then we can evaluate the amount of angular
momentum that has been injected into the halo. In Fig.~\ref{figAM}
we show the specific averaged angular momentum profiles as yielded
by the old, NFW-like (\textit{dashed}), and new (\textit{solid})
DFs, in the same potential. Keep in mind that this is not a stable
state, the system is going to evolve into a configuration that
satisfies the Poisson equation.

The final configuration of the system, corresponding to its new
equilibrium state, is showed in Fig.~\ref{figNEW}. The density
profile (\textit{upper left}) has been smoothed and the cusp erased;
this mirrors the mass loss by the inner regions of the halo, since
the DM particles have been moved to more energetic orbits. Notice
the corresponding flattening of the potential well (\textit{upper
right}). Accordingly, the rotation curve (\textit{inset}) is more
gently rising in the inner regions and is nearly unchanged beyond
half of the virial radius.

As to the velocity dispersion, the symmetry between the radial and
tangential motions has been broken in favor of the latter
(\textit{lower left}); relatedly, the anisotropy parameter profile
(\textit{lower right}) is now negative in the inner halo and changes
sign at a radius corresponding to the new $r_a=1/\sqrt{2}$, that is
set by requiring the same outermost value of $\beta(r)$ as before,
although it scarcely affects the results shown above.

In sum, given the DF of equation (\ref{fnostra}), the couple ($\alpha=0, \
r_a \sim 1$) describes the known NFW halo, cuspy and almost entirely
isotropic, while the couple ($\alpha=1/2, \ r_a \sim 1/\sqrt{2}$)
produces a corelike halo, with a constant inner
density profile and a tangentially-dominated inner anisotropy
profile.

We point out that the injection of angular momentum corresponding to
our DF produces additional bulk rotational motions, though they are
not to be intended as \textit{ordinate} in the common sense of the
word; in fact, the symmetry of the halo and its dynamics are defined
by that of the DF, and therefore there is not a definite
preferential direction of rotation. The DM particles move on orbits
around the center of mass with random orientations, so that the
resulting angular momentum $L^2$ is nonzero, but the average value
of each velocity component is null.

\section{Dynamical friction as an angular momentum engine}

Now the issue arises: Is there a physical process that can account
for the evolution described above? In this section we aim at
presenting a toy model that provides the halo with such an amount of
angular momentum to allow its DF to evolve from a cuspy to a cored
configuration. The natural framework is galaxy formation, when the
baryons collapse inside the dark halo potential well and exchange
angular momentum with the DM through dynamical friction.

Specifically, in the very early stages of galaxy formation, the
baryons trapped inside the potential wells of the halos undergo
radiative dissipation processes that cause them to lose kinetic
energy and to form clumps inside the relatively smooth dark halo. If
radiative cooling is effective, the gas will organize into
self-gravitating clouds before it collapses to the halo center and
condenses into stars; the clouds are likely to survive the tidal
stripping due to the DM, because of their relatively high binding
energy (Mo \& Mao 2004). The clumpy gas component decouples from
virial equilibrium, and dissipates its orbital energy; in detail,
the clouds spiralling down get closer and closer to the halo center,
increasing their tangential velocity along their orbits and reaching
regions with higher and higher density.

In these regions, a gravitational effect becomes relatively
efficient in slowing down the clouds, namely the dynamical friction
exerted by the background DM particles, that causes part of the
cloud velocity to be transferred from the baryons to the DM itself (El-Zant et
al., 2001, 2004).
As a result, the inner part of the halo is granted with a surplus of
angular momentum and energy, depending on the number, mass and
initial velocity of the clouds.

Consider a cloud of mass $M_c$ that at time $t=0$ is at a certain
distance from the center of the halo, with initial velocity
$v^2=v^2_r+(L/r)^2$ and angular momentum $L$; we define the initial
pericenter of its orbit as $r_+(0)$, the eccentricity as $e(0)$, and
the apocenter as $r_-(0)= r_+\, (1-e)/(1+e)$. The cloud is
self-gravitating and hence we consider it as a point mass, immersed
in the halo potential well; in the orbit-averaged approximation
(Lacey \& Cole 1993), the equations of motions for the cloud energy
and angular momentum are given by
\begin{equation}
{\mathrm{d} E\over \mathrm{d} t}  = - {\int_{r_-}^{r_+}{(1/v_r)}~~
v\, |F_{\mathrm{frc}}| / M_c \ \mathrm{d}r \over
\int_{r_-}^{r_+}{(1/v_r)} \ \mathrm{d}r}~, \label{derLdyn}
\end{equation}
\begin{equation}
{\mathrm{d} L\over \mathrm{d} t}  = - {\int_{r_-}^{r_+}{(1/v_r)}~~
L\, |F_{\mathrm{frc}}| / (M_c v) \ \mathrm{d}r \over
\int_{r_-}^{r_+}{(1/v_r)} \ \mathrm{d}r}~, \label{derEdyn}
\end{equation}
with initial conditions set by
\begin{equation}
L(0)=\sqrt{2[\Psi(r_+)-\Psi(r_-)]\over
1/r_+^2-1/r_-^2}~,~~~~~~~~~~~E(0)=\Psi(r_+)+\frac{v^2(r_+)}{2}~.
\label{LE}
\end{equation}
At each instant, the force exerted by the background DM particles on
the cloud is
\begin{equation}
|F_{\mathrm{frc}}| =-4\pi G^2\, M_c^2\, \ln{\Lambda}\ {\int_0^v
f(v')~{\mathrm{d}^3v'}\over v^2}~, \label{Fdyn}
\end{equation}
in terms of the NFW phase-space distribution function $f$ (see
Section 2), of the cloud speed $v=\sqrt{2[\Psi(r)-E(t)]}$ and of the
Coulomb logarithm $\ln{\Lambda} = \ln{(M_H/M_c)}$. At each timestep,
$r_\pm(t)$ are given by the condition $v_r^2=\sqrt{v^2-L^2/ r^2}=0$.
Due to the dynamical friction, the orbit shape and the velocity of
the cloud evolve in time, so that this set of equations has to be
solved iteratively.

For a halo of virial mass $M_H$ we performed a series of Montecarlo
simulations for different ensembles of clouds, characterized by a
mass function scaling as $M_c^{-\delta}$, with index $\delta$
ranging from $0$ to $2$; in each realization, we allowed the cloud
masses to range from $10^{-5}$ to $10^{-2}\, M_H$ (El-Zant et al.
2004). The number of clouds is actually constrained by the total
amount of baryons, set to equal the cosmological fraction $0.16\,
M_H$. The initial spatial distribution of the clouds is uniform
between $r=0$ and $r=R_H$; at time $t=0$ the clouds are in
statistical equilibrium with the background halo, therefore their
initial velocities are randomly sampled from a Maxwellian
distribution, with mean $0$ and variance $\langle \sigma^2_{t, r}
\rangle/2$.

For each $\delta$ we performed $100$ runs, and computed the average
specific angular momentum transferred by the clouds to the halo
after $2$ Gyr; we expect that after this time the inner part of the
halo becomes so crowded with clouds that they start to collide and
disrupt, and the star formation effects dominates over the dynamics
of the gas (El-Zant et al. 2004). However, in the outer regions the
process continues with longer timescales, so that we also followed
the evolution of the system for about $6$ Gyr.

\begin{deluxetable}{lccccc}
\tabletypesize{} \tablecaption{Dynamical friction results}
\tablewidth{0pt} \tablehead{\colhead{$\delta^a$} & \colhead{$\langle
N_i \rangle^b$} & \colhead{$\langle M_c \rangle^c $} &
\colhead{$\langle \Delta L_{DM} \rangle^d$} & \colhead{$M_g$ ($2$
Gyr)$^e$} & \colhead{$M_g$ (6 Gyr)$^f$}} \startdata
0 & 31.45 & 5.03e-3 & 1.04 & 0.039 & 0.074  \\
1 & 109.66 & 1.46e-3 & 1.86 & 0.032 & 0.064 \\
2 & 2276.98 & 7.05e-5 & 7.54 & 0.019 & 0.043 \\
\enddata
\label{clouds} \tablecomments{Column label: (\textit{a}) cloud
power-law mass function index; (\textit{b}) average number of
clouds; (\textit{c}) average cloud mass (units of $M_H$);
(\textit{d}) average of the exchanged specific angular momentum
(units of $R_H\ V_H$) integrated over the profile; (\textit{e})
average baryonic mass that falls inside $0.1\, R_H$ after $2$ Gyr
and, (\textit{f}) after $6$ Gyr (units of $M_H$).}
\end{deluxetable}

\begin{figure}\epsscale{0.5}\plotone{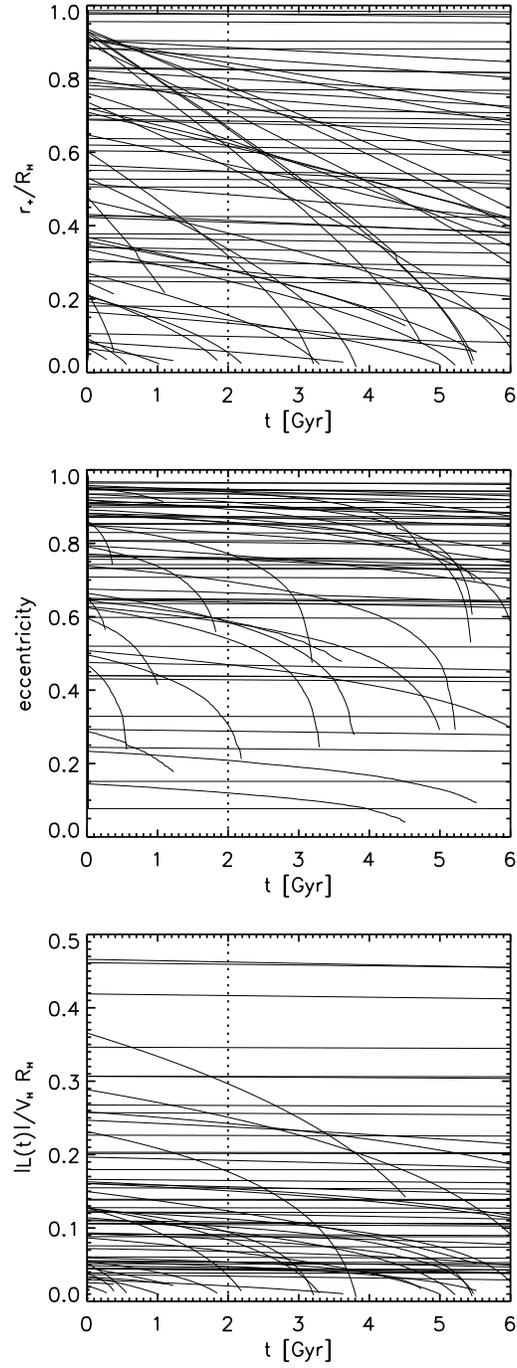}\caption{Time evolution of
a set of clouds sampled with power-law index $\delta=1$.
\textit{Upper panel}: apocenter of the orbit; \textit{middle panel}:
orbit eccentricity; \textit{bottom panel}: angular momentum retained
by the clouds.} \label{figclouds}
\end{figure}

In Table~\ref{clouds}, for each $\delta$ we give the average number
and mass of the clouds, the total angular momentum gained by the
halo, and the mass accumulated in the center of the halo after $2$
and $6$ Gyr. Note that with our power-law mass functions, the
massive clouds constitute a small fraction of the total; on the
other hand, the dynamical friction is more effective on them, and
therefore they have a large probability to lose all their angular
momentum quickly, and to collapse in the center of the halo soon.
The small clouds instead take more time to spiral down in the halo
potential well, and retain a larger fraction of their initial
angular momentum. In the end, a steeper mass function, that selects
a high number of small clouds, results in a more effective transfer
of angular momentum to the halo.

\begin{figure*}\plotone{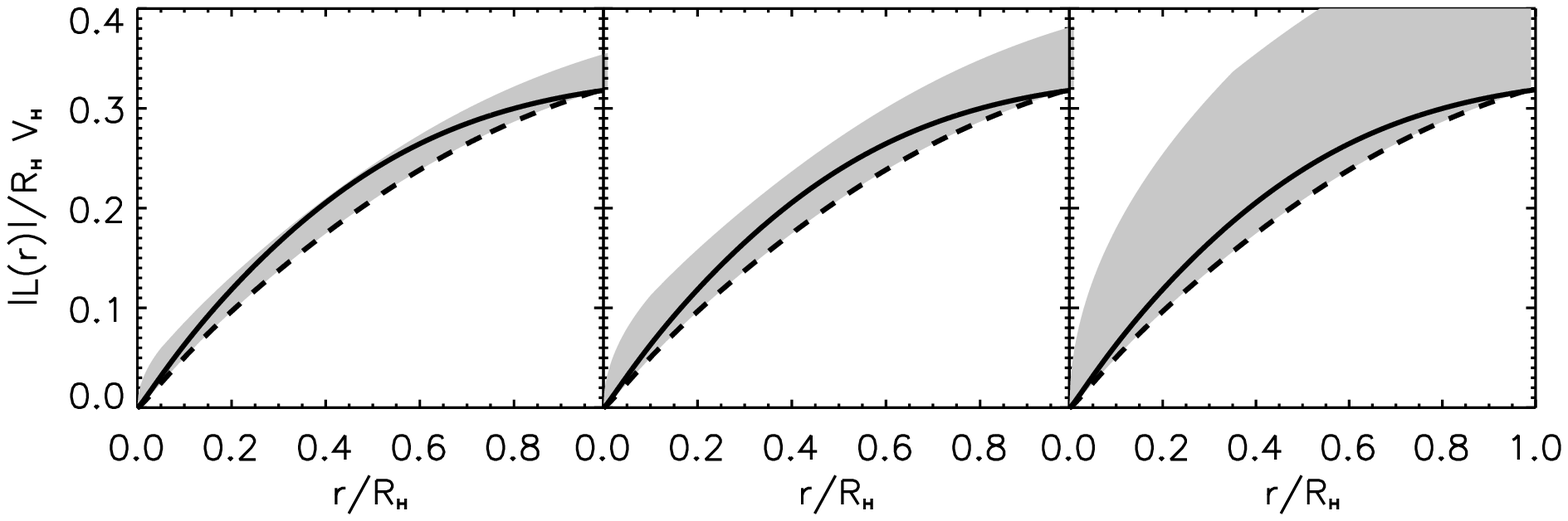}\caption{Comparison between
the angular momentum profiles implied by the phase-space DFs and by
dynamical friction. \textit{Dashed line}: unperturbed NFW halo
(see Section 2); \textit{solid line}: perturbed halo, new DF (see
Section 3); \textit{shaded regions}: NFW angular momentum profile
plus $\Delta L$ from dynamical friction, for $\delta=0$
(\textit{left}), $\delta=1$ (\textit{center}) and $\delta=2$
(\textit{right}).} \label{figAM3}
\end{figure*}

In Fig.~\ref{figclouds} we illustrate an example of the time
evolution of a set of clouds sampled with power-law index
$\delta=1$. In the time interval from $t=0$ to $t=6$ Gyr a fraction
of the clouds reaches the inner $10\%$ of the virial radius (upper
panel), building up the mass that is likely to end up in the
spheroidal component of the forming galaxy. In the middle panel we
show the evolution of the orbit eccentricity, and in the lower panel
the angular momentum lost by the single clouds and transferred to
the halo. Notice that the fraction of clouds that effectively
transfer angular momentum to the halo is relatively small, and that
the majority of the clouds remains on almost unperturbed high
orbits, meaning that the timescales of dynamical friction are long.
This effect is enhanced for increasing $\delta$, as shown in the
last two columns of Table~\ref{clouds}; a steeper mass function
results in a slower accumulation of baryons in the center. In any
case, the final amount of baryons inside the inner $10\%$ of the
virial radius after $2$ and $6$ Gyr is overabundant with respect to
the known galactic masses (spheroidal and/or disk components; see
Shankar et al. 2006), due to the fact that the total baryonic
component initially matches the cosmological fraction. However,
while baryons accumulate in the center of the potential well and
organize themselves into the protogalactic structure, star formation
and AGN activity start, along with the ensuing feedback processes
that eventually regulate the actual amount of baryons.

Notice that these feedback mechanisms can in no way affect the
transfer of angular momentum between the visible and dark
components, as they take place after the cloud collapse; as will be
discussed in the next section, the growing baryonic mass at the
center of the halo could affect the final dynamical state of the
system only marginally.

In Fig.~\ref{figAM3} we present the final angular momentum profile
of the halo, for $\delta=0, 1, 2$. From each run of our simulation
we extract a transferred momentum profile after $2$ Gyr and then add
it to the original NFW (\textit{dashed lines}, see also
Fig.~\ref{figAM}); the \textit{shaded areas} represents the overlap
of all the resulting new profiles. For comparison, we plot the
profile obtained in Section 3 through the transformation of the DF
(\textit{solid lines}, see also Fig.~\ref{figAM}).

\smallskip

The angular momentum profile produced through the dynamical friction
mechanism is clearly compatible with that resulting from the
perturbation of the halo DF described in Section 3. In fact, the
effect of the dynamical friction on the halo is that of enhancing
the tangential motions with respect to the radial; this is the same
behavior expected in a halo whose DF is a function of $L^2$,
regardless of the details in the shape of the DF itself.

\section{Discussion and conclusions}

We approached the topic of the coevolution of the baryons and their
host DM halo, to investigate whether the mechanism of galaxy
formation can flatten the inner halo density profile and reconcile
the observational evidences with the standard theory of hierarchical
clustering.

We performed two independent computations over the NFW halo, one
perturbing its phase-space distribution function with angular
momentum, the other following the evolution of the baryonic angular
momentum during the collapse inside the halo potential well; we are
faced with the conclusion that both of them are viable to describe
the halo evolution due to galaxy formation, the former from the
microscopic and the latter from the macroscopic point of view.

We showed, through the study of the microscopic dynamical properties
of the DM, that the NFW halo is strongly sensitive to angular
momentum transfer; an injection of angular momentum to the DM
produces a flattening of the density profile and a redistribution of
the particles in velocity space, with an increase of the fraction of
tangential motions over the total velocity dispersion.

From the macroscopic side, we looked for a mechanism of angular
momentum transfer that directly involved the baryons, that was
especially efficient during galaxy formation, and that directly
affected the microscopic state of the system; thus we focused on
dynamical friction (El-Zant et al., 2001, 2004). Notice that such a process 
is well below the
current resolution limit of the cosmological numerical simulations; however, 
it is implemented in recent simulations of cluster and galactic size halos 
(Nipoti et al. 2004, Ma \& Boylan-Kolchin 2004). In fact,
angular momentum can be transferred also by tidal interactions and
mergers; however, such events are less and less frequent with
decreasing redshift both for spheroidal galaxies (Koopmans et al.
2006, Dominguez-Tenreiro et al. 2006) and for spirals (D'Onghia \&
Burkert 2004, Tonini et al. 2006). Moreover, such interactions between the
halo and its neighbours occur on scales comparable to that of the halo
itself, thus giving rise to global perturbations of its dynamical state and 
global ordinate motions; for
this reason, it is more likely that such events affect the halo spin rather 
than the halo inner angolar momentum and anisotropy profiles.

When modeling the phase-space DF, we had to face the degeneracy
intrinsic to this kind of problem, that allows infinite functions to
satisfy the Poisson equation. The symmetry of the system is set by
that of the DF; for an isotropic system, the only integral of motion
is energy, and $f$ is uniquely determined. But if the symmetry
between the radial and tangential velocity components is broken,
then $f$ becomes $f(\varepsilon, L^2)$; the system is bound to
conserve the internal angular momentum, and when perturbed with a
$\Delta L^2$, it is enforced to find a new equilibrium
configuration. This is the case we considered in Section 3; we
focused on a power-law in $L^2$ due to its simplicity, and showed
that it can account for the basic observables of the halo, such as
the density and anisotropy profiles, and the shape of the potential.
Note that this kind of DF cannot account for the halo spin, that is
found to be $\lambda \sim 0.03-0.06$ from simulations and
observations (see Tonini et al. 2006 and references therein). This
value of $\lambda$ is small, nonetheless it gives us information on
the true shape of the halo in phase-space. In fact, more generally,
the system would be described by a DF of the form $f(\varepsilon,
L^2, L_z)$, where also the symmetry between the two tangential
velocity components is broken. A complete shape of the perturbed DF
could be of the kind $f(Q,L^2,L_z)=f_0(Q) \, f_1(L^2, L_z)$, where
$f_1$ may be a power-law or a more complex function, odd in $L_z$ to
have net bulk rotation around the symmetry axis. With this DF the
tangential velocity $\vec{v_T}$ is parallel to the equatorial plane
and the halo features a macroscopic total angular momentum in the
$z$-direction, $\vec{L_z} = \vec{r} \times \vec{v_T}$, that is
aligned with the spin, $\lambda \simeq v_T/\sigma$.

In the computation of the perturbed gravitational potential we did
not include the baryonic mass, although it does have an effect on
the total gravitational potential, since the baryons piling up in
the center of the halo tend to deepen the well. But three
considerations are needed here: 1) the right amount of baryons that
is actually observed in the center of halos is deeply connected with
feedback processes, that are responsible for removing at least half
of the initial baryonic mass (Shankar et al. 2006); 2) as hinted in
Section 1, the feedback processes themselves can transfer energy to
the DM and contribute to smoothen the cusp; 3) the dominant effect
in changing the equilibrium configuration of the halo is the DF
symmetry breaking induced by the angular momentum transfer, while
the deepening of the potential well simply produces an isotropic
enhancement of all the $\sigma$ components, that does not interfere
with the described process. So, even taking into account the
deepening of the total potential well due to baryons, the effect
turns out to be quite negligible.

Up to now, there is no definite knowledge of the details of the
baryon collapse into the protogalactic structure. Nevertheless, if
dynamical friction indeed plays a major role in the collapse, it
could affect the morphology of the galaxy that is to form, possibly
through its timescale that in turn depends on the cloud mass
function. While massive clouds, collapsing early and losing all
their momentum, can feed up the spheroidal component, the small
clouds are most likely to end up in a rotating disk; in particular,
the slow, gradual accretion of small clouds can be assimilated to
the infall of material that causes the \textit{inside-out} formation
of disks (Chiappini et al. 1997). As a possible evidence of this
process, we can consider the molecular clouds that today travel
across the Galactic disk, trapped inside the potential well of the
DM halo; these are too small to have actively participated in the
growth of the disk, because only marginally affected by dynamical
friction.

\medskip

The main results of the paper are the following:

$\bullet$ we computed the NFW phase-space distribution function, and
analyzed the internal halo dynamics, with particular focus on the
anisotropy and the angular momentum profiles.

$\bullet$ we perturbed the NFW halo by injecting angular momentum
into it, \textit{i.e.} breaking the symmetry of its DF that becomes
explicitly a function of the two variables $E$, $L^2$; physically
speaking, the DM particles gain orbital energy and bulk tangential
motions. As a consequence, the halo evolves into a new equilibrium
configuration, characterized by a cored density profile, a
tangential anisotropy in the inner regions, and a shallower
potential well; we also computed the new angular momentum profile.

$\bullet$ we tested that dynamical friction exerted by the
DM on the baryons at the time of protogalaxy formation is a viable
mechanism to perturb the halo structure, a result in agreement with El-Zant et
al. (2001, 2004). In particular we show that the
transfer of angular momentum from the baryons to the DM due to dynamical friction
produces an angular momentum profile consistent with the one
of a halo that has been perturbed and eventually features a tangential
velocity anisotropy and a corelike profile.

We consider this evolutionary approach most promising in solving the
cusp-core controversy that affects the interpretation of the
observational data, and in moving towards the understanding of the
mechanisms of galaxy formation.

\begin{acknowledgements}
We thank the referee for her/his comments and suggestions. 
Work supported by grants from ASI, INAF and MIUR.
\end{acknowledgements}

\end{document}